\documentclass{article}

\renewcommand{\vec}[1]{\overrightarrow{#1}}

\usepackage{graphics}
\usepackage{graphicx}
\usepackage{amssymb}
\usepackage{array}
\usepackage{fancyhdr}
\usepackage{subfigure}
\usepackage{graphicx}
\usepackage{mathrsfs}
\usepackage{slashed}
\usepackage{stmaryrd}
\usepackage{epsfig}
\usepackage{dcolumn}
\usepackage{bm}
\usepackage{amsmath}
\usepackage{wrapfig}
\usepackage{makeidx}
\usepackage{wasysym}
 \usepackage{relsize}
\usepackage{la}
\usepackage[utf8]{inputenc}
\usepackage{listings}
\usepackage{color}

\usepackage{array}
\usepackage{fancyhdr}
\usepackage{graphicx}
\usepackage{mathrsfs}
\usepackage{slashed}
\usepackage{stmaryrd}

\begin{document}

\title{\bf Cosmological Constant in  the Imaginary-Time Field Theory}
\author{Yi-Cheng Huang\\
E-mail: {\tt ychuang1109@msn.com}
}
\date{}
%
%
\maketitle
{\bf Abstract}
\vskip.2cm
{
The origins of the cosmological constant are discussed from the perspective of the imaginary-time field theory. The concept of the thermal time, which is related to the Tolman-Ehrehfest relation, and the conformal invariance of the actions are applied to account for the relation between the scale factor of the FRW metric and the temperature of the vacuum. Finite values of the cosmological constant from the DeWitt-Schwinger representation and the Casimir effect with a large separation between two plates  are derived. The induced energy density is found to be uniform over the space and independent of the evolution of the universe, and the equation of state ratio is indeed $w=-1$. From the point of view presented here, the largest discrepancy of the vacuum energy between the theoretical and the experimental sides can be conciliated.  And the value of the cosmological constant corresponds to 
a characteristic temperature of vacuum determined by the history of the universe. 
 %
%

%
\newpage
\section{Introduction}
\label{intro}
 The analogy between the thermodynamics and the general relativity is gradually explored and  under disputes for decades. 
In 1915, the gravitation was successfully described by Einstein \cite{einstein15} through the geometrization of the space and time. It was not taking long later, many experimental tests for the theory of relativity have helped to achieve its authenticity and excellence in understanding the weakest force in the universe. No more than two decades, in 1930, Tolman et al. \cite{tolman30}  examined the thermodynamical properties of a perfect fluid and  radiation in generating  gravitational fields through the Einstein equation. The Tolman-Ehrenfest relation states the relation between the temperature and the metric. Then it started in  the 1960s and 1970s, the general relativity can be further understood in the thermodynamical perspective through studying the collapse of the black hole. In the theory constructed by Hawking et al. \cite{hawking73}, the dynamics of black holes are described by the temperature, entropy and so on, which are all familiar terminologies in the thermodynamics. It is found that a black hole emits black-body radiation near the event horizon with a characteristic temperature, the so-called Hawking radiation. 
It was about the same time, Fulling, Davies and Unruh \cite{unruh} published respectively a prediction of  a thermal radiation that would be detected by an accelerated observer, and the thermal bath was depicted by the Unruh temperature.    
It is believed that there is a deep connection  between the Hawking radiation and the Unruh temperature since they are equivalent if the equivalence principle is applicable on them. Even though the effects of the Hawking radiation and Unruh temperature are hard to detect, for instance, a temperature of $1$K corresponds to a proper acceleration of $\sim10^{21}\,m/s^2$, it was shown recently that a well known effect, called the  Sokolov–Ternov effect \cite{steffect}, in the experiments of accelerator physics, is in fact the Unruh effect under certain conditions.  Besides, the notion of the thermal time was introduced by  Connes and Rovelli \cite{rovelli93}  in the 1990s. A thermal time flow is argued to be determined by any thermal state in the covariant quantum theory, and so as to define the physical time. Besides, this concept can also be proved to agree with the Tolman-Ehrenfest relation. In short, the existence of the thermodynamical characteristics in the general relativity can not be denied, and it is worth as many attentions as in other approaches, like loop gravity,  etc., in order for a theory of quantum gravity to come. \par
On the other hand, both in the theoretical and experimental sides, the appearance of the cosmological constant continues to puzzle physicists for nearly one century. It was first conjectured by Einstein \cite{einstein17} in 1917 to construct a static universe as well as to satisfy the Mach principle \cite{mach} for a connection between the mass density of the universe and the geometry. Soon after Hubble discovered the expansion of the universe, Einstein \cite{pais82} retreated the new additional term and called it his greatest blunder. However, from the studies of the modern cosmology, it turns out that it is mainly responsible for the cosmic acceleration at the present time and taking up $\sim 70\%$ composition of  the universe from the observations of  Type Ia supernova \cite{iasupernova} and the cosmological background radiation \cite{cmb}. Nevertheless, the cosmological constant is still no more than mysteries to physicists. Its energy density is constant over the space regardless of the expansion of the universe, and the pressure that it exerts on  the universe is negative with an equation of state ratio $w=-1$.  In the cosmology it is often regarded as the vacuum energy, while  in the quantum field theory the vacuum energy is yielded through summing over zero-point energies of all normal modes of some fields up to an energy cutoff  of the Planck scale, $M_{\rm Pl}\sim 10^{18}$ GeV. The estimation  from the theory and the observations gives the greatest discrepancy of a physical quantity in physics, a difference of $\sim10^{120}$ orders of magnitude \cite{weinberg89}.
\par
In the imaginary-time field theory \cite{huang13a}, a temperature-dependent scale transformation is introduced to the actions of the quantum electrodynamics for both of photons and fermions and the imaginary-time formalism gives a thermodynamical nature to the vacuum, which happens to coincide the thermodynamical perspective of the general relativity. Not only the ultraviolet divergences can be removed from the theory itself but one-loop radiative corrections of the QED are proved to be in agreement with those in the conventional field theory, besides it predicts the same renormalization group equations as those in the $\overline{MS}$ renormalization scheme \cite{msrenormalization}. Moreover, in ref. \cite{huang13b}, the imaginary-time field is found to generate  consistent results with the known vacuum effects, such as the Casimir effect and the van der Waals force \cite{vanderwaals}, and shows deep connections with the Unruh effect and the Hawking radiation. The goals of this paper are to relate the thermodynamical features in the general relativity with the newly developed imaginary-time field theory  and attempts to investigate possible origins  of the cosmological constant from an integrated aspect. In the following section, the concept of the thermal time is briefly described and its relation with the imaginary-time is discussed, including applications on the Rindler coordinates and the FRW metric. In the section \ref{dewittschwinger}, the calculation of the cosmological constant with the DeWitt-Schwinger's approach is presented, the induced vacuum energy is proved to possess the features of being a cosmological constant. Then the derivations of the Casimir effect for calculating the vacuum energy of the electromagnetic waves and fermions are adopted in Section \ref{casimir}; the discrepancy of the $120$ orders of magnitude can be found to be  conciliated. In the end, a conclusion is attached.

\section{Conformal Invariance in spatially flat spacetime}
\label{conformal}

\subsection{Thermal time and scale invariance}
\label{thermal}

The connection between the general relativity and the thermodynamics was investigated in the early stage through the Tolman-Ehrenfest effect \cite{tolman30}, which states the relation
between the temperature of a statistical system, $T$, in a gravitational field and time component of the metric:   
\begin{eqnarray}
T\sqrt{g_{00}}=const.\,.\label{TErelation}
\end{eqnarray}
The perfect fluid and the radiation were being examined through the respective Einstein equation to yield this   property in common. In the Newtonian limit, the above relation can be reduced to 
\begin{eqnarray*}
\frac{1}{T}\vec{\nabla} T&=& \vec{g},
\end{eqnarray*}
where $\vec{g}$ is the Galilean acceleration of gravity. It was concluded that an increase in equilibrium temperature was found to accompany a decrease in gravitational potential. The concept of the thermal time \cite{rovelli93} developed in the 1990s was inspired by the above effect, and was proved to work in the covariant quantum theory. The idea is briefly described as follows. For observables $A$ in the Poisson algebra $\mathcal{A}$ over a phase space $\mathcal{S}$. 
Given the state of a system, $\rho$, in $\mathcal{S}$, the  thermal time, $t$, is defined by the time flow, $\alpha^\rho_t: \mathcal{A}\rightarrow \mathcal{A}$, as the Poisson flow of $-\ln \rho$ in $\mathcal{A}$:
\begin{eqnarray*}
\frac{d\alpha^\rho_t(A)}{dt}&=&-\left\{A,\ln\rho\right\}.
\end{eqnarray*}
We may also define the Newtonian mechanical time $\tau$ for the time flow of the observable $A$ as
\begin{eqnarray*}
\frac{dA}{d\tau}&=&\left\{A,H\right\},
\end{eqnarray*}
where $H$ is the hamiltonian of the system. As a non-relativistic Boltzman-Gibbs equilibrium state $\rho_T$ reaches a equilibrium temperature $T$, the state can be expressed as 
\begin{eqnarray*}
\rho_T&\propto&e^{-\frac{H}{k_BT}}.
\end{eqnarray*}
The relation of the thermal time, $t$, and the mechanical time, $\tau$, which is identified as the proper time, can be found as
\begin{eqnarray}
\frac{d}{dt}&=&\frac{1}{k_B T}\frac{d}{d\tau}.\label{thermalT1}
\end{eqnarray}
\begin{figure}[t]
\begin{center}
   \subfigure[]{\includegraphics[width=5.9cm]{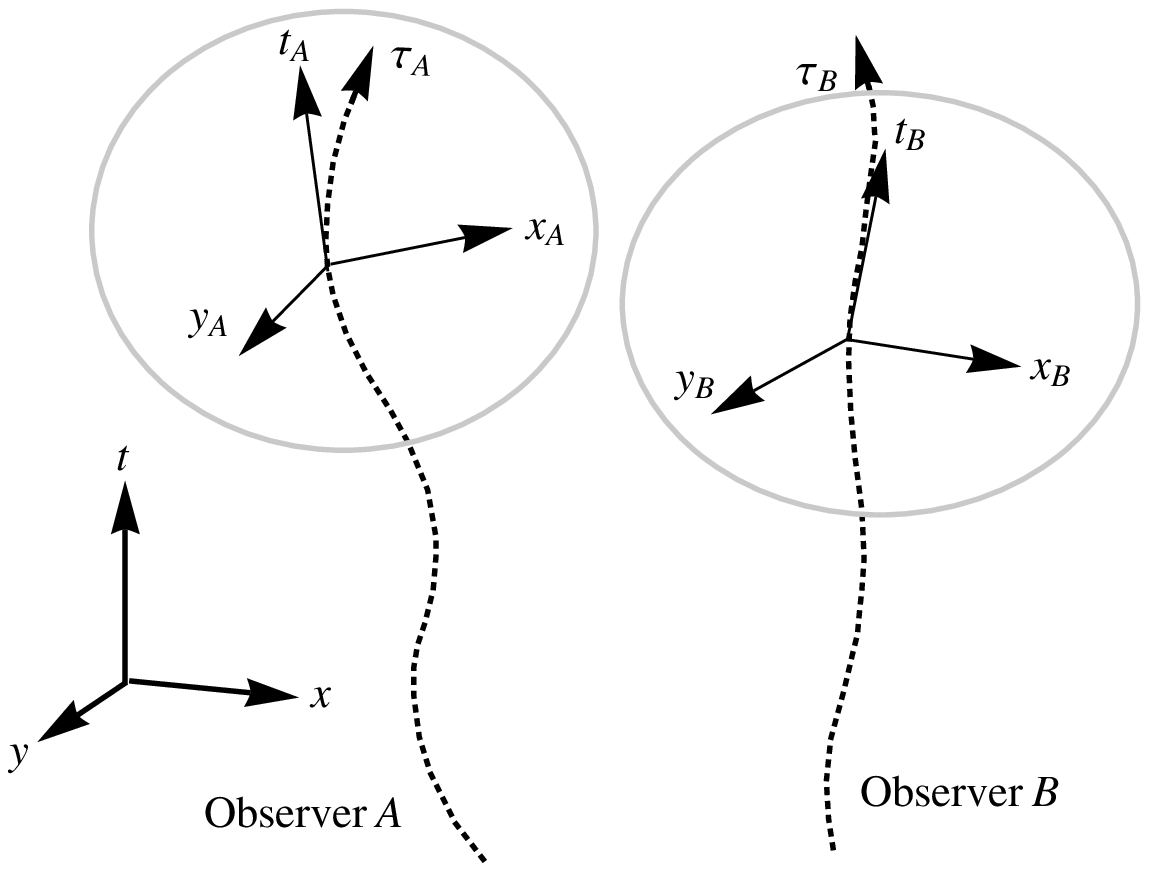}}
   \hspace*{0.00\textwidth}
   \subfigure[]{\includegraphics[width=5.9cm]{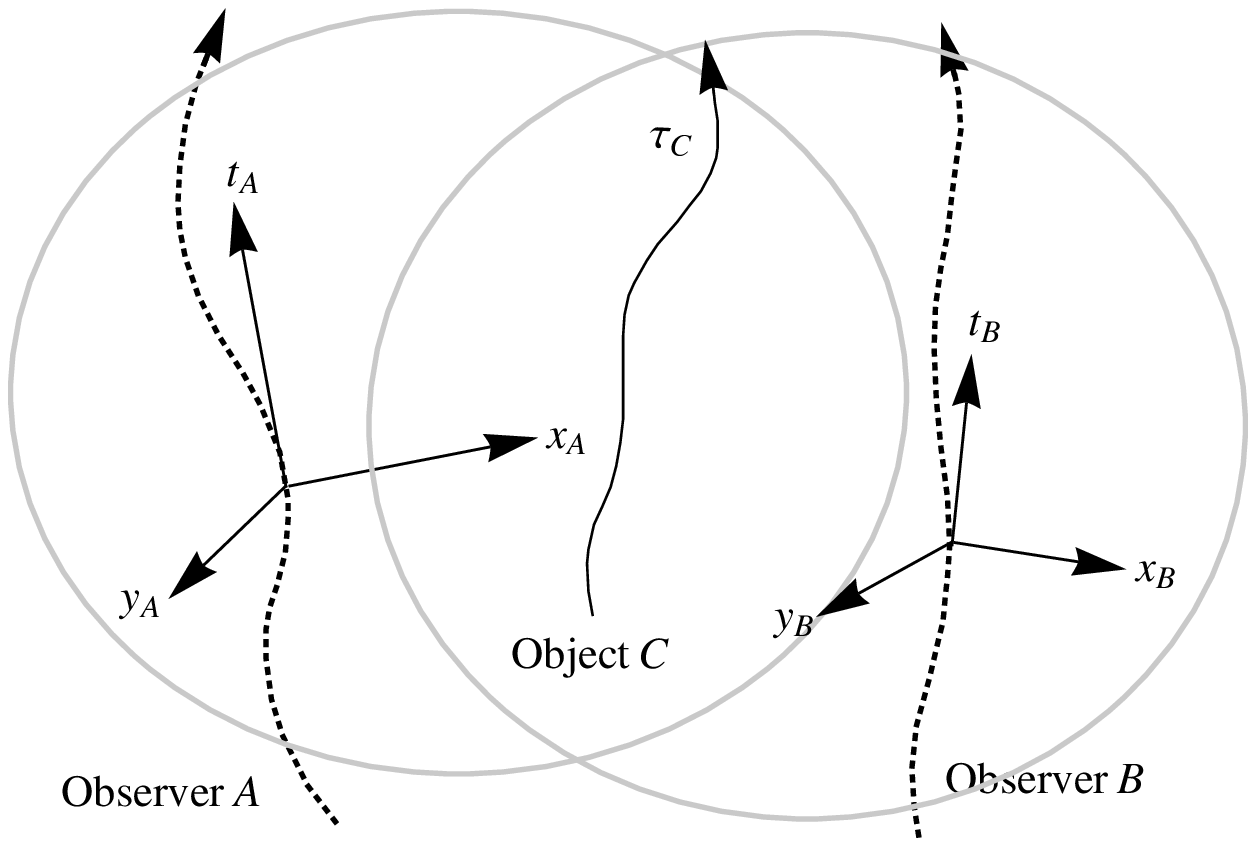}}
   \hspace*{0.0\textwidth}
\caption{\small  Suppose that in a global 3-dimensional space-time, $(t,x,y)$ there are two observers $A$ and $B$,  whose world lines are drawn as two dotted lines.  For the observer A, at a proper time $\tau_A$ he or she can build a local inertia frame $(t_A,x_A,y_A)$ in the neighborhood to describe physical laws, and of course $t_A=\tau_A$. The same work can be done by the observer $B$. The gray circles indicate their respective local frames.} 
  \label{Fig:thermalT}
\end{center}
\end{figure}
The thermal time $t$ is a global variable over the whole space-time and the mechanical time, or say the proper time $\tau$, depends on distinct observers. For example, imagine there are two observers $A$ and $B$ moving in a 3-dimensional space-time, $(t,x,y)$, as illustrated in Figure \ref{Fig:thermalT} (a). For each of the observers, he or she can construct a local inertial frame: $(t_\mathsmaller{I},x_\mathsmaller{I},y_\mathsmaller{I})|_{(\mathsmaller{ I}=A,B)}$, along the world line with the corresponding proper time $\tau_\mathsmaller{I}$, such that $\tau_\mathsmaller{ I}=t_\mathsmaller{ I}$. From eq. (\ref{thermalT1}), we may conclude that, for the two observers' measurements on time,  $\frac{t_A}{\beta_A}=\frac{t_B}{\beta_B}$, for both of them are associated with the same global time $t$. In addition, we may consider this concept in another viewpoint. If  there is an object $C$ with  a proper time $\tau_\mathsmaller{\rm C}$ which is measured by two observers $A$ and $B$ in their respective local frames as shown in Figure \ref{Fig:thermalT} (b), it has the Tolman-Ehrenfest Relation: $d\tau_\mathsmaller{\rm C}={C_{T.E.}}{\beta_\mathsmaller{\rm C}}dt$, where $C_{\rm T.E.}$ is the coefficient from the Tolman-Ehrenfest relation. Then it can be related to the observers' time  by $d\tau_\mathsmaller{\rm C}=\frac{\beta_\mathsmaller{\rm C}}{\beta_\mathsmaller{\rm I}}dt_\mathsmaller{\rm I}$ (I=A,B). After rescaling out the redundant factor $\beta_\mathsmaller{\rm C}$, we may reach 
\begin{eqnarray}
t_\mathsmaller{\rm I}&=&\beta_\mathsmaller{\rm I} \tau_\mathsmaller{\rm C}, \hspace{.3cm}{\rm where}\,\, \beta_\mathsmaller{\rm I}=\frac{1}{k_BT_\mathsmaller{\rm I}}, \label{tautrelation}
\end{eqnarray} 
for an arbitrary observer ${\rm I}$. Now it may look easier to understand the relation, $\frac{t_A}{\beta_A}=\frac{t_B}{\beta_B}$.
Further discussions can be found in ref. \cite{rovelli93} for interested readers. Based from the above discussion, in ref. \cite{menicucci11} it  can  be further argued that for two observers A and B located at different constant positions, or say in two subregions of the system, each watch measures the global time $t$ in terms of the respective proper time through the formula $d\tau_\mathsmaller{I}=\sqrt{g_{00}}dt$, and from the Tolman-Ehrenfest relation the relation between the proper times is $T_Ad\tau_A=T_Bd\tau_B$. In the local inertial coordinate frames used by the observer A and B,  $d\tau_\mathsmaller{I}= \eta_{ij}dx^idx^j$, where $\eta_{ij}={\rm diag}(1,-1,-1,-1)$, therefore the above relation can extend to both of time and spatial coordinates for the physics described by two observers in separate regions.
From the above discussions, it can be deduced that for temperatures in different subregions of a system the local times at different locations are related with each other through $\frac{t_A}{\beta_A}=\frac{t_B}{\beta_B}$ and $\frac{{\bf x}_A}{\beta_{A}}=\frac{{\bf x}_B}{\beta_B}$. In ref. \cite{huang13a}, both of the actions for photons and fermions are required to have the scale invariance, the scale transformation for the time variable is $t_A=\frac{\beta_A}{\beta_B}t_B$, which coincides with the above discussion for the thermal time, as well as the transformation for the three spatial dimensions, ${\bf x}_A=\frac{\beta_A}{\beta_B}{\bf x}_B$. Besides, in the field theory of the curved space-time, the conformal transformation is often used on the field equations and is thus in a deep connection with this idea; more related explanations will be presented in the next subsection. 
This concept can be further understood from the perspective of the renormalization group. In Wilson's approach \cite{wilson75} in the condensed matter physics, the scale change on the Lagrangian can lead to effects, like running couplings, etc.\,.
Those effects can be discovered in the calculations of  one-loop radiative corrections of ref. \cite{huang13a}, as the actions of fermions and photons are invariant under the temperature-dependent scale transformations, and naturally any physical phenomenon caused by the scale changes is dependent on the variation of the temperature. Thus there is no wonder that the coefficients of renormalization group equations in the imaginary-time  field theory  are found to be identical to those of the $\overline{MS}$ renormalization scheme in the field theory and  the temperature  plays the role of the energy scale factor $\mu$. It is noticed that these individual concepts, which were once scattered in different branches of physics, can be reasoned in an integrated way now. In the following, an example of the Rindler's metric \cite{rindler} for the thermal time will be shown to have a connection with the Unruh temperature, and the idea will be applied to the FRW metric \cite{frw} for the cosmological model. It will help us to interpret the results from the calculations of the cosmological constant that is performed in Section \ref{dewittschwinger} and \ref{casimir}.

\subsection{An application for a uniformly accelerated observer}
\label{example}
The Rindler coordinate system \cite{rindler} is the frame of reference for a uniformly  accelerated observer in the Minkowski space.  We may take a two dimensional Mikowski space $(t,x)$ with the metric, $ds^2=dt^2-dx^2$, for example without losing any generality. The Rindler's coordinate described by $(\eta,\xi)$ with a constant, $a$, is related to the Minkowski space by the transformation
\begin{eqnarray}
&&\hspace{-1.cm}t=a^{-1}e^{a\xi}\sinh a{\eta}, \hspace{.2cm}x=a^{-1}e^{a\xi}\cosh a\eta,\,{\rm with\,\, the\,\, metric}\,\,\,\,ds^2=e^{2a\xi}(d\eta^2-d\xi^2).\hspace{.5cm}\label{rindlermetric}
\end{eqnarray}
The Rindler's chart of $(\eta,\xi)$ only takes up one fourth of the Minkowski space $(t,x)$ for $x>|t|$, and is also called Rindler wedge. The proper acceleration is defined as $\alpha^\mu=(\frac{du^0}{ds},\frac{du^1}{ds})$, where the 4-velocity is $u^\mu=(\frac{dt}{ds},\frac{dx}{ds})$, then the norm of the proper acceleration is $\alpha=\sqrt{-\alpha^\mu \alpha_\mu}=ae^{-a\xi}$. For the study of quantum field theory in the Rindler space \cite{birrel}, the metric in eq. (\ref{rindlermetric}) is conformal to the whole Minkowski space, for the conformal transformation $g_{\mu\nu}\rightarrow e^{-2a\xi}g_{\mu\nu}$, therefore the wave equation is conformal invariant in the Rindler space. By matching the quantum fields between the vacuum states in the Minkowski space, $|0_M\rangle$, and those in the Rindler space $|0_R\rangle$ through the Bogoliubov transformations \cite{bogoliubov,birrel}, it can be found that for the creation and annihilation operators, $b^\dagger_{\bf k}$, $b_{\bf k}$, of a Rindler observer, he or she will detect  a thermal bath of the vacuum state $|0_M\rangle$
\begin{eqnarray*}
\langle 0_M|{b_{\bf k}}^\dagger b_{\bf k}|0_M\rangle&=&\frac{1}{e^{2\pi\omega/a}-1},\,{\rm where}\,\, \omega=|{\bf k}|.
\end{eqnarray*}  
This is the spectrum of a black-body radiation with a temperature $T_0=\frac{a}{2\pi k_B}$, and it is also called the Unruh temperature. As for the accelerated observer, the temperature is given by the Tolman-Ehrenfest relation in eq. (\ref{TErelation}), now the constant can be set to be 
\begin{eqnarray}
T\sqrt{g_{00}}=T_0,\,\label{TErelation2}
\end{eqnarray}
where $\sqrt{g_{00}}=e^{a\xi}$ from the metric in eq. (\ref{rindlermetric}). We may obtain the Unruh temperature $T=\frac{\alpha}{2\pi k_B}$, where $\alpha$ is indeed the proper acceleration. More detail about the above discussion can be found in ref. \cite{birrel}. Now we may turn the attention to the current model of the universe, the Fiedman-Robertson-Walker metric (FRW) \cite{frw}:
\begin{eqnarray}
ds^2=dt^2-a^2(t)\left(\frac{dr^2}{1-kr^2}+r^2d\Omega^2\right),\label{frw}
\end{eqnarray}
where $k$ is the constant curvature of the space. We will assume the universe a flat space for $k=0$ and introduce the conformal time, $\eta$, for the FRW's universe,
\begin{eqnarray}
t=\int^t dt'=\int^\eta a(\eta')d\eta',
\end{eqnarray}
therefore the metric in eq. (\ref{frw}) becomes
\begin{eqnarray}
ds^2=a^2(\eta)\left(d\eta^2-{dr^2}{}-r^2d\Omega^2\right),\label{cfrw}
\end{eqnarray}
where $a^2(\eta)$ is the conformal scale factor. From the discussion in the beginning of this section, the conformal time, $\eta$, is identified as the thermal time and the temperature of the vacuum is determined by 
\begin{eqnarray}
T\,a(\eta)=const.\,,\label{Ta}
\end{eqnarray}
where the temperature, $T(\eta)$,  is also a function of $\eta$. The constant can be set to be the current temperature $T_0$ at the present time $\eta_0$ so that $a(\eta_0)=1$. The function $a(\eta)$ is the scale factor at the time $\eta$.  Therefore, the eq. (\ref{Ta}) simply tells us that the scale factor is the ratio of the temperatures in different eras. We can denote it as $a(\eta)=\frac{\beta(\eta)}{\beta_0}$.  Like the discussion for the accelerated observer, quantum fields can be quantized  for the vacuum states in the current Minkowski space, $|0_{a(\eta_0)}\rangle$, and those in the Minkowski  space in another era, $|0_{a(\eta)}\rangle$. If the physics observed by local observers of distinct periods is believed not different from one another, the actions which defines the Lagrangian densities should be independent of the expansion of the universe, in other words, should be scale invariant.  And it is known that the actions of fermions and photons discussed in \cite{huang13a} are invariant under such temperature-dependent scale transformations. In Section \ref{dewittschwinger}, the cosmological constant deduced by the DeWitt-Schwinger representation is calculated in the imaginary-time field theory  and the scale factor considered here will be used to account for the reason why the energy density is a constant over all of the space and through the evolution of time.



\section{DeWitt-Schwinger representation}
\label{dewittschwinger}
The action in the general relativity can be expressed as $S=S_g+S_m$, where the gravitational action is 
\begin{eqnarray}
S_g=\int \frac{d^d x}{16\pi G}\sqrt{-g}\left(R-\Lambda\right),\label{Sg}
\end{eqnarray}
and $S_m$ is the action of the matter.
The classical Einstein equation is derived by the condition $\frac{\partial S}{\partial g_{\mu\nu}}=0$. The variation on the scalar curvature, $R$, gives rise to the  deformation of the space and time, and the cosmological constant, $\Lambda$, is needed to explain the acceleration of the universe from the cosmological observation data. In the context of the field theory, the DeWitt-Schwinger representation \cite{schwinger51,dewitt75} is able to give a prediction for the cosmological constant. Unfortunately, a traditional problem that occurs in field theories also happens, the predicted value is divergent \cite{birrel}, and a bare quantity is also needed in order to fit the experimental data. As  the formalism of the imaginary-time field theory \cite{huang13a} is known to be free of UV divergences, it would be naturally to see if it still can give a finite answer to the cosmological constant. In obtaining the Green functions in curved spacetime \cite{schwinger51,dewitt75,bunch79}, the Riemann normal coordinate $y^\mu$ for a point $x$ is adopted to expand the metric tensor as well as the Green functions. The notations presented here are followed by those in ref. \cite{birrel}. The idea is briefly described as follows. In general, the generating function of a field $\phi$ can be obtained after finishing a Gaussian integral of $\phi(x)$ and the corresponding source function $J(x)$,
\begin{eqnarray*}
Z[J]&\propto&\left[\det(-G_F)\right]^\mathsmaller{\frac{1}{2}}\exp\left[-{\frac{i}{2}}\int d^dx d^dx' J(x)G_F(x,x')J(x')\right],
\end{eqnarray*} 
where the prefactor can be rewritten as 
$
\left[\det(-G_F)\right]^\mathsmaller{\frac{1}{2}}=\exp\left[\mathsmaller{\frac{1}{2}}{\rm tr} \ln(-G_F)\right]
$.
An effective action can be defined as a function, $W\equiv \ln Z[0]$ as well as the corresponding effective Lagrangian
\begin{eqnarray}
W&=&\int d^dx \mathscr{L}_{\rm eff}(x),\,\,\, {\rm where}\,\,\mathscr{L}_{\rm eff}={ \frac{1}{2}}{\rm tr}\ln\left(-G_F\right).\label{effaction}
\end{eqnarray} 
The trace of the operator $\ln\left(-G_F\right)$ can be computed through
\begin{eqnarray}
{\rm tr}\ln \left(-G_F\right)&=&\int d^dx \sqrt{-g(x)}\lim_{x\rightarrow x'}\langle x |\ln\left(- G_F\right)|x'\rangle.\label{lnGf}
\end{eqnarray} 
And a useful integral representation that is used in the precedent calculations is 
\begin{eqnarray}
\frac{1}{k^2-m^2+i\epsilon}&=&-i\int ds \,e^{is(k^2-m^2+i\epsilon)}.\label{intrep}
\end{eqnarray} 
The resultant effective Lagrangian $\mathscr{L}_{\rm eff}$ contributes to the cosmological constant in eq. (\ref{Sg}), but is divergent. The term of the leading order is like 
\begin{eqnarray}
-\frac{4m^2}{(4\pi)^{d/2}d(d-2)}\left\{\frac{1}{d-4}+\frac{1}{2}\left[\gamma_\mathsmaller{\rm E}+\ln\left(\frac{m^2}{\mu^2}\right)\right]\right\}+...\,,\label{Lcl}
\end{eqnarray} 
where $d$ is the dimension of the space-time. 
The remainder is proportional to the scalar curvature $R$ and is ignored here since it is beyond the scope of this paper.

\subsection{Cosmological constant from Electromagnetic waves}
\label{emwaves}
In the imaginary-time formalism of field theory \cite{huang13a}, the Matsubara frequency plays the role of the energy for the imaginary-time dimension. Take the metric tensor $g_{\mu\nu}$ aside without losing any generality for a second and will restore it in the end, the imaginary-time Green function of photons can be written as
\begin{eqnarray*}
\hspace{-.5cm}G_F(\omega_n,{\bf q})=\frac{1}{\omega_n^2+{\bf q}^2}
,&{\rm and}&G_F(x,x')=\frac{1}{\beta}\sum_n\int\frac{d^3{\bf q}}{(2\pi)^\mathsmaller{{3}{}}}e^{-i\omega_n(\tau-\tau')+i{\bf q}\cdot (x-x')}G_F(\omega_n,{\bf q}),
\end{eqnarray*} 
where the metric tensor $g_{\mu\nu}$ is ignored for a simplicity reason and will be restored in the end. In order not to be confused with a factor $\beta$ that will be introduced later for a temperature at some time in the past, the factor $\beta=\frac{1}{k_BT}$ will be set to $\beta_0=\frac{1}{k_BT_0}$ from now on in this section for it to represent  the temperature of vacuum at the present time.
In practice, similar to eq. (\ref{intrep}) for the imaginary-time Green function of photons the integral representations that will be used are
\begin{eqnarray}
\hspace{-.8cm}\frac{1}{i\omega_n-|{\bf q}|}
=-\int^{\beta_0}_0 ds\,\, \frac{e^{{\beta_0}|{\bf q}| }}{e^{{\beta_0} |{\bf q}|}-1}e^{(i\omega_n-|{\bf q}|)s},&{\rm and}&\,\,
\frac{1}{i\omega_n+|{\bf q}|}
=\int^{\beta_0}_0 ds\,\, \frac{1}{e^{{\beta_0} |{\bf q}|}-1}e^{(i\omega_n+|{\bf q}|)s}.\nonumber\\
\label{formula3}
\end{eqnarray}
In order to take the integration, we may introduce another integral $S(m)$, and it contains the exponential function ${\rm E_1}(...)$ with a parameter $m$, which will be set to zero later. The lower bound of the integration domain, $\epsilon$, is an infinitesimal positive number.
\begin{eqnarray}
S(m)&\equiv&\int^\infty_\epsilon \frac{ds}{s} e^{-i(\omega_n^2+{\bf q}^2+m )s}={\rm E_1} \left(i(\omega_n^2+{\bf q}^2+m )\epsilon\right) 
\,\,\nonumber\\&=&-\gamma_\mathsmaller{\rm E} -i\mathsmaller{\frac{\pi}{2}}-\ln(\omega_n^2+{\bf q}^2+m)-\ln \epsilon+O(\epsilon),\label{sa}
\end{eqnarray}
where $\gamma_\mathsmaller{\rm E}$ is the Euler-Gamma constant and $m$ has a dimension of $[{\rm mass}]^2$. By taking the derivative of the above formula, we may reach the following relations
\begin{eqnarray*}
\hspace{-.5cm}\frac{d S(m)}{d m}=-i\int^\infty_0 {ds} \,\,e^{-i(\omega_n^2+{\bf q}^2+m )s}=-\frac{1}{\omega_n^2+{\bf q}^2+m},\,\, {\rm and}{\hspace{.3cm}}
S(m)=-\int^\infty_m dm' \frac{d S(m')}{d m'}.
\end{eqnarray*}
From eq. (\ref{sa}), if the remainder function $O(\epsilon)$ is neglected, we may obtain the formula
\begin{eqnarray*}
\langle \omega_n,{\bf q}|\ln \left(-G_F\right)| \omega_n,{\bf q}\rangle=\ln \left(\frac{-1}{\omega_n^2+{\bf q}^2}\right)=S(0)+\gamma_\mathsmaller{\rm E}+i\mathsmaller{\frac{3\pi}{2}}
+\ln\epsilon.
\end{eqnarray*}
The logarithm function appears to be divergent after integrating over the 3-momenta and summing over  the frequency, so it is hoped that the formalism of the  imaginary-time  field theory will bring a regularization function to make a finite answer, just like what we had in the calculations of the Casimir effect and the  van der Waals force \cite{huang13b}. However, from the above expression, the last three constants apparently will lead to a divergence. If a finite result is possible from the right-hand side of the above, the integration for the function $S(0)$ has to cancel the divergence.  Based from this approach, the calculation of the function $\langle x|\ln (-G_F)|x'\rangle$ is separated into two parts, $I_1$ and $I_2$; notations $\Delta {\bf x}={\bf x}-{\bf x'}$ and $\Delta\tau=\tau-\tau'$ will be used and set to zero or near zero in the following.   It will soon be realized that the cancellation will be made possible by choosing an infinitesimal value of $\Delta \tau$.  For the first part, 
\begin{eqnarray*}
I_1&=&\frac{1}{{\beta_0}}\sum_n\int\frac{d^3{\bf q}}{(2\pi)^3}S(0)e^{-i\omega_n\Delta\tau+i{\bf q\cdot \Delta x}}\nonumber\\
&=&
-\int^\infty_0 dm \frac{1}{{\beta_0}}\sum_n\int\frac{d^3{\bf q}}{(2\pi)^3}\frac{d S(m)}{d m} e^{-i\omega_n\Delta\tau+i{\bf q\cdot \Delta x}}\\
&=&\int^\infty_0 dm \frac{1}{{\beta_0}}\sum_n\int\frac{d^3{\bf q}}{(2\pi)^3}\frac{1}{\omega_n^2+{\bf q}^2+m}e^{-i\omega_n\Delta\tau+i{\bf q\cdot\Delta x}},
\end{eqnarray*}
where the factor $\frac{1}{\omega_n^2+{\bf q}^2+m}$ is expanded into $\frac{1}{2\sqrt{{\bf q}^2+m}}\left(\frac{1}{i\omega_n+\sqrt{{\bf q}^2+m}}-\frac{1}{i\omega_n-\sqrt{{\bf q}^2+m}}\right)$.
While setting ${\bf \Delta x}=0$ and temporarily keeping $\Delta \tau$ as nonzero, apply the formulas in eq. (\ref{formula3}) then we obtain
\begin{eqnarray*}
&&\hspace{-1.5cm}I_1=\int^\infty_0 dm \int^{\beta_0}_0 ds \frac{1}{{\beta_0}}\sum_n\int\frac{ |{\bf q}|d|{\bf q}|}{2\pi^2}\left(\frac{e^{-\sqrt{|{\bf q}|^2+m}s}}{2}+\frac{\cosh(\sqrt{|{\bf q}|^2+m}s)}{e^{{\beta_0} \sqrt{|{\bf q}|^2+a}}-1}\right)e^{-i\omega_n(\Delta \tau-s) }\frac{{|{\bf q}|}}{\sqrt{{\bf q}^2+m}},\\
&=&\int^\infty_0 dm  \int\frac{ |{\bf q}|d|{\bf q}|}{2\pi^2}\left(\frac{e^{-\sqrt{|{\bf q}|^2+m}\Delta \tau}}{2}+\frac{\cosh(\sqrt{|{\bf q}|^2+m}\Delta \tau)}{e^{{\beta_0} \sqrt{|{\bf q}|^2+m}}-1}\right)\frac{{|{\bf q}|}}{\sqrt{{\bf q}^2+m}}.
\end{eqnarray*}
The sum of $e^{-i\omega_n(\Delta\tau-s)}$ over the Matsubara frequencies in the first line yields a delta function, $\delta(s-\Delta\tau)$, which can be integrated out by $\int^{\beta_0}_0ds$ right away. The above integration converges as long as $\Delta \tau$ is kept finite, and the multiple integral of $m$ and $|{\bf q}|$ will be combined into one single integral.  Let $m=k^2_1+k^2_2$ and $|{\bf q}|^2=k^2_3+k^2_4$, so that $dm=\frac{1}{\pi}dk_1dk_2$ and $|{\bf q}|d|{\bf q}|=\frac{1}{2\pi}dk_3dk_4$. Combine them together $dm |{\bf q}|d|{\bf q}| =\frac{1}{2\pi^2}d^4{\bf k}$. We can make a variable change to a four-dimensional hyper-spherical coordinate $(k,\theta_1,\theta_2,\theta_3)$ for $(k_1,k_2,k_3,k_4)$:
\begin{eqnarray*}
\hspace{-.5cm}k_4=k\sin\theta_3\sin\theta_2\sin\theta_1,\,\,\,k_3=k\cos\theta_3\sin\theta_2\sin\theta_1,\,\,\, k_2=k\cos\theta_2\sin\theta_1,\,{\rm and}\,\,k_1=k\cos\theta_1,
\end{eqnarray*}
where $0<\theta_1<2\pi$ and $0<\theta_2,\theta_3<\pi$. The volume element in the new coordinate is  $\frac{1}{2\pi^2}d^4{\bf k}=\frac{k^3}{2\pi^2}dkd\Omega$, where $k=\sqrt{k_1^2+k_2^2+k^2_3+k^2_4}=\sqrt{m+|{\bf q}|^2}$ and the solid angle $d\Omega=\sin^2\theta_3\sin\theta_2d\theta_3d\theta_2d\theta_1$. The last factor in the above integration can be rewritten as $\frac{{|{\bf q}|}}{\sqrt{{\bf q}^2+m}}=\frac{\sqrt{k^2_3+k^2_4}}{k}=|\sin\theta_2\sin\theta_1|$.
The result of the first part of the integration becomes 
\begin{eqnarray}
&&\hspace{-1.3cm}I_1={\frac{1}{4\pi^4}}\int^\infty_0 k^3 dk\left(\frac{e^{-k\Delta \tau}}{2}+\frac{\cosh(k\Delta \tau)}{e^{{\beta_0} k}-1}\right)\int^{2\pi}_0d\theta_1\int^\pi_0d\theta_2d\theta_3\sin^2\theta_3\sin\theta_2|\sin\theta_2\sin\theta_1|\nonumber\\
&=&
{\frac{1}{4\pi^2}}\left(\frac{3}{(\Delta \tau)^4}+\frac{\pi^4}{15{\beta_0}^4}\right).\label{db1}
\end{eqnarray}
As mentioned before, the first term in the parenthesis blows up as $\Delta \tau$ goes to zero, and we need the cancellation of the divergence in the second part. The second part of the integration, $I_2$, for  ${\bf \Delta x}=0$ and $\Delta \tau=0$
are 
\begin{eqnarray}
I_2=(\gamma_c+\ln \epsilon)\frac{1}{{\beta_0}}\sum_n\int\frac{d^3{\bf q}}{(2\pi)^3}=(\gamma_c+\ln \epsilon)\frac{\Lambda^4_{\rm cutoff}}{(2\pi)^4}=\ln (e^{\gamma_c} \epsilon)\frac{\Lambda^4_{\rm cutoff}}{(2\pi)^4},
\label{gaep}
\end{eqnarray}
where, similar to ref. \cite{huang13a}, the cutoff of the Matsubara frequency is chosen to equal to that of the 3-momentum, $N_{\rm max}=\frac{\Lambda_{\rm cutoff}}{2\pi}{\beta_0}$, and $\gamma_c=\gamma_\mathsmaller{\rm E}+i\mathsmaller{\frac{3\pi}{2}}$. As $\epsilon\rightarrow 0$, the integration, $I_2$ turns to be negative.  We may choose $\Delta \tau$ to cancel the divergences in eq. (\ref{db1}) and (\ref{gaep})
\begin{eqnarray}
\ln (e^{\gamma_c} \epsilon)\frac{\Lambda^4_{\rm cutoff}}{(2\pi)^4}+\frac{3}{4\pi^2\Delta \tau^4}=0\hspace{.5cm}&\Rightarrow &\hspace{.5cm}\Delta \tau=
\frac{\frac{\beta_0}{N_{\rm max}}}{\sqrt[4]{\frac{4\pi^2}{3}\ln\left(\frac{1}{e^{\gamma_c}\epsilon}\right)}}.\label{tauconditionDS}
\end{eqnarray}
Therefore for the cutoff $\Lambda_{\rm cutoff}\rightarrow \infty$ and $\epsilon\rightarrow 0$, an infinitesimal  value of $\Delta \tau$ can be picked to satisfy the cancellation. The above condition for $\Delta\tau$ is similar to the condition in eq. (\ref{taucondition}) in the next section; it implies that $\Delta\tau$ is the smallest distance that can be resolved by photons.  There will be more discussions about its physical meaning when we get to the vacuum energy for the Casimir effect.  Taking into account the trace of $g_{\mu\nu}$ and the prefactor $\frac{1}{2}$ in eq. (\ref{effaction}), the finite part in eq. (\ref{db1}) contributes to the effective Lagrangian density, also the energy density of virtual photons, $\bar{\varepsilon}^{\rm D.S.}_{0,\gamma}$, in eq. (\ref{lnGf}) is 
\begin{eqnarray}
\mathscr{L}_{\rm eff}&=&\bar{\varepsilon}^{\rm D.S.}_{0,\gamma}
=\frac{\pi^2}{30{\beta_0}^4}.
\label{dsconst}
\end{eqnarray}
The Stephen-Boltzmann constant of the black-body radiation is $\sigma=\frac{\pi^2}{60}$. Unlike the divergent result in eq. (\ref{Lcl})  obtained in the conventional framework, the prediction given by the 
imaginary-time field theory is finite as expected in the beginning of the paper. The problem now is how to interpret it and relate it to the cosmological constant that we know. As we know from the modern cosmology, the energy density of the cosmic background radiation (CMB) is proportional to $1/a^4$, where $a$ is the scale factor of the universe, due to the expansion of the universe and the extension of their wavelength. In Section \ref{conformal}, we know from the Tolman-Ehrenfest relation the relation between the temperature and the scale factor, $a(\eta)=\frac{\beta(\eta)}{\beta_0}$. As a result, the energy density in the current age of the universe for photons is
\begin{eqnarray}
\bar{\varepsilon}^{\rm D.S.}_\gamma(\eta_*)=\frac{\bar{\varepsilon}_{0,\gamma}^{\rm D.S.}}{a^4}=\frac{\pi^2/(30\beta_0^4)}{\beta_*^4/\beta_0^4}=\frac{\pi^2}{30{\beta_*}^4},
\end{eqnarray}
where $\eta_*$ is some time in the evolution of the universe for the initial value of the   energy density, $\bar{\varepsilon}_{\gamma}^{\rm D.S.} $. This implies that a characteristic temperature, which we may denote it as $\beta_*=\beta(\eta_*)$, was recorded in the electromagnetic vacuum during the history of the universe, similar to the CMB. When the CMB temperature cools down to about $3000$ K, it signifies the beginning of the recombination era and its evolution separates from other elements in the universe. On the other hand, the energy density  of the vacuum stays the same no matter how the scale factor changes along $\beta_0$. Its difference from the CMB is that the temperature of the CMB does not correlate with the scale factor, so the temperature of the CMB declines as the universe expands. The current observation of the cosmological constant is $\sim 10^{-47}$ GeV$^4$, if we assume that the electromagnetic wave is the only element in the vacuum, then the corresponding characteristic temperature is $\beta_*\sim 27.3$ K. 




\section{Casimir effect }
\label{casimir}					
As mentioned before, the observational data of the cosmological constant is about $\sim 10^{-47}\, {\rm GeV}^4$. With the ordinary quantum field theory up to the Planck scale $M_{\rm Pl}=(8\pi G)^{-1}\sim 10^{18} $ ${\rm GeV}$, the zero-point energies of all normal modes of some fields give rise to a vacuum energy $\sim 10^{72}$ ${\rm GeV}^4$, which leads to the famous discrepancy of $120$ orders of magnitude between the experimentally observed and the theoretically predicted values of the cosmological constant. Obviously, what behind this huge disagreement should be some weaknesses in our theoretical formalism. In the previous work \cite{huang13b} on the applications of the imaginary-time field theory on various vacuum effects, the calculation of the Casimir effect based from the new proposed theory was found to agree with those in the classical approach.
Hence, it would be interesting to see if the same theory could tell anything new about the vacuum energy. The experimental setup of the Casimir effect is  that two parallel conducting plates, each is a square of a length $L$, are placed with a distance $d$ between them, and the energy per unit area, which is stored in between, will be calculated. We may adopt the same mathematical approaches that have been used in ref. \cite{huang13b} for the electromagnetic waves and the fermions in the following derivations.

\subsection{Energy density and negative pressure of vacuum }
The Hamiltonian of the electromagnetic waves based on the imaginary-time field theory is 
\begin{eqnarray}
\mathcal{H}_0(\omega_n,{\bf q})
&=&\frac{1}{4|{\bf q}|}(\omega_n^2-|{\bf q}|^2)\sum_\lambda (-g_{\lambda\lambda})
\left(a^\lambda_{\omega_n,{\bf q}} a^{\lambda\dagger}_{\omega_n,{\bf q}} + a^{\lambda\dagger}_{\omega_n,{\bf q}} a^\lambda_{\omega_n,{\bf q}}\right),\label{h0}
\end{eqnarray}
where $\lambda$ is the polarization index and $a^{\dagger}_{\omega_n,{\bf q}} $ and $ a^{\lambda\dagger}_{\omega_n,{\bf q}} $ are annihilation and creation operators of the photon field. The matrix $g_{\lambda\lambda'}={\rm diag}(1,-1,-1,-1)$, which is not a tensor, simply expresses the sign of each polarization state. The average energy of the system after summing over all of the Matsubara frequency is 
\begin{eqnarray}
\langle  \mathcal{H}_0({\bf q})\rangle
=\lim_{\tau\rightarrow 0^+}|{\bf q}|\left({e^{-|{\bf q}|\tau}}{}+2\cosh (|{\bf q}|\tau) n_B(|{\bf q}|)\right).\label{aveh0w}
\end{eqnarray}
Consider the experimental setup of the Casimir effect, two plates are placed in parallel to the $x$-$y$ plane, and the distance $d$ is along the $z$-direction.  The energy stored between two plates is $E(d)$, so the energy per unit area of the conducting plate is   
\begin{eqnarray*}
\frac{E(d)}{L^2}=\lim_{\tau\rightarrow 0^+}\frac{1}{\pi^2}{\sum^\infty_{n=0}} \,'\int^\infty_0 dq_x \int^\infty_0 dq_y|{\bf q}|\left({e^{-|{\bf q}|\tau}}{}+2\cosh (|{\bf q}|\tau) n_B(|{\bf q}|)\right),
\end{eqnarray*}
where $|{\bf q}|=\left(q^2_x+q^2_y+\frac{n^2 \pi^2}{d^2}\right)^\mathsmaller{1/2}$ and $n$ is an integer. The summation ${\sum_n}'$ indicates that an extra  factor $1/2$ is inserted for $n=0$. Remember in ref. \cite{huang13b}, the second term in the parenthesis is dropped due to the large value of ${\beta_0}$ in the density function $n_B(|{\bf q}|)$. In the case considered here, the distance $d$ between the two plates will be taken to infinity and the factor, ${\beta_0}$, is kept as a large but finite value, and the integration from the second term will be taken into account. The summation and the integration of the energy density will be separated into two parts, for each term in the parenthesis, $\bar{\varepsilon}=\bar{\varepsilon}_1+\bar{\varepsilon}_2$. As for the first,
\begin{eqnarray}
\hspace{-1cm}\bar{\varepsilon}_1&=&\lim_{{\tau\rightarrow 0^+}}\frac{1}{\pi^2d}{\sum^\infty_{n=0}} \,'\int^\infty_0 dq_x \int^\infty_0 dq_y|{\bf q}|{e^{-|{\bf q}|\tau}}{}=\lim_{{\tau\rightarrow 0^+}}\frac{1}{4\pi }\left(\frac{12}{\pi\tau^4}-\frac{\pi^3}{180\,d^4}+O(\tau^4)\right).\nonumber\\
\label{aveE}
\end{eqnarray}
\begin{figure}[t]
\begin{center}
   \subfigure[]{\includegraphics[width=5.8cm]{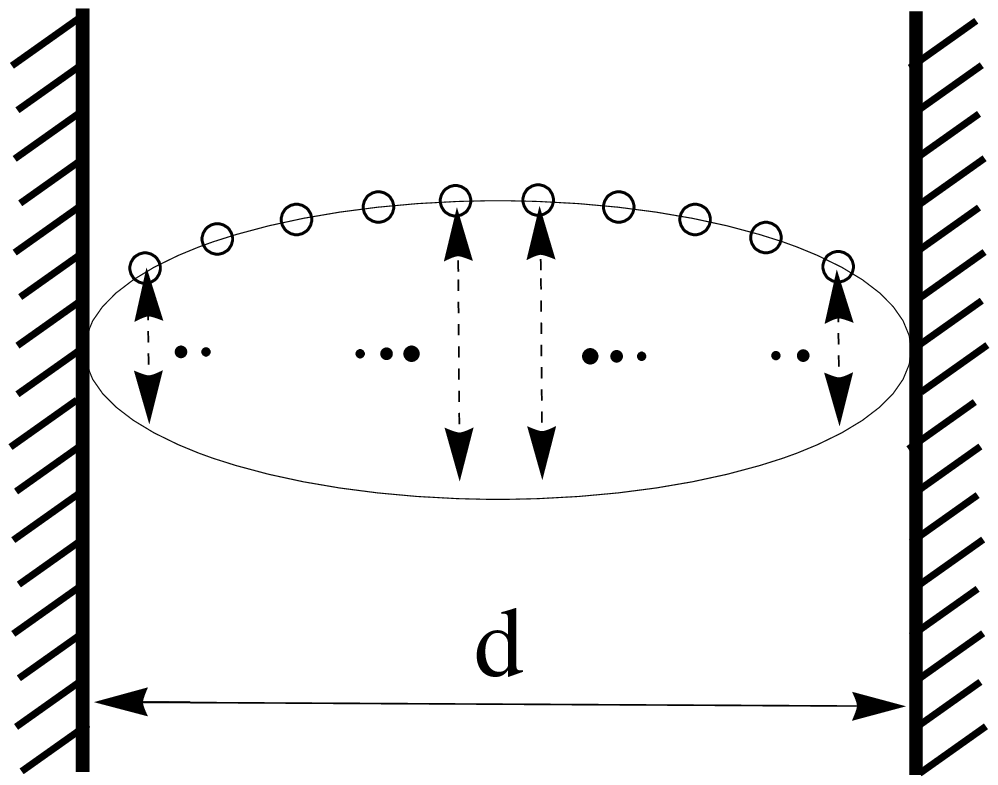}}
   \hspace*{0.01\textwidth}
   \subfigure[]{\includegraphics[width=5.8cm]{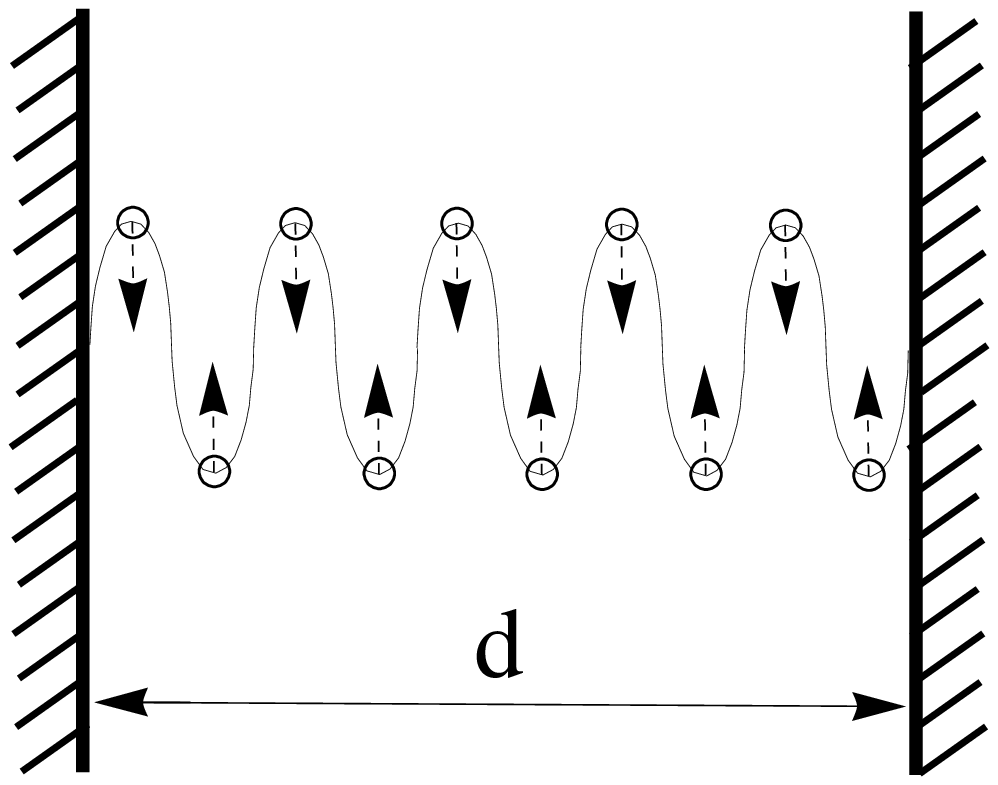}}
   \hspace*{0.0\textwidth}
\caption{\small Suppose there are $N=10$ particles in the experimental setup of the Casimir effect as shown in Fig. (a). The behaviors of the particles in two parallel walls are described by the harmonic oscillating motion, and each particle is a harmonic oscillator. In Fig. (b), the standing waves formed by the macroscopic behavior of all of the oscillators have a largest mode number $N_{\rm max}=N$, and it corresponds to a maximal frequency that carried by the waves, $f_{\rm max}=\frac{N_{\rm max}}{2d}c$, where $c$ is the speed of the wave. More importantly, it has a shortest wavelength $2d/N_{\rm max}$, which is similar to the condition of the characteristic length $\tau$ in eq.  (\ref{taucondition}). Therefore the factor $e^{-|{\bf q}|\tau}$ appeared in the average energy  of eq. (\ref{aveh0w}) is  simply to impose a constraint on the highest energy that a single wave can reach in a system based on the above reason.   } 
  \label{Fig:cutoffgraph}
\end{center}
\end{figure}
Similar to the discussion in \cite{milonni,huang13b}, the divergent term of $1/\tau^4$ is neglected, only the term of $1/d^4$ makes a solid contribution. As for a small distance $d$, the traditional result of the induced Casimir force for the electromagnetic waves is restored. We did not care the divergent term, since it does not depend on $d$ as we measure the Casimir force with varying distances. However, if a finite value is expected for the vacuum energy, we have to see what brings a divergence to our derivation. Let's take one step backward in the calculation and see what might be the reason. 
We may give the integration of $q_x$ and $q_y$  a cutoff ${\Lambda_{\rm cutoff}}$, instead of  an infinity in eq.  (\ref{aveE}), then restart. The first integral becomes
\begin{eqnarray}
\bar{\varepsilon}_1&=&\lim_{{\tau\rightarrow 0^+}}\frac{1}{\pi^2d}{\sum^\infty_{n=0}} \,'\int^{\Lambda_{\rm cutoff}}_0 dq_x \int^{\Lambda_{\rm cutoff}}_0 dq_y|{\bf q}|{e^{-|{\bf q}|\tau}}{}\nonumber\\
&=&\lim_{{\tau\rightarrow 0^+}}\frac{1}{2\pi d}{\sum^\infty_{n=0}} \,'\int^{\Lambda_{\rm cutoff}}_0 q_\perp dq_\perp |{\bf q}|{e^{-|{\bf q}|\tau}}{},
\end{eqnarray}
where $q_\perp=\sqrt{q_x^2+q_y^2}$ and $|{\bf q}|=\sqrt{q^2_\perp+\frac{n^2\pi^2}{d^2}}$. Then we have 
\begin{eqnarray}
\bar{\varepsilon}_1&=&\lim_{{\tau\rightarrow 0^+}}\frac{1}{2\pi d}{\sum^\infty_{n=0}} \,'\int^{\Lambda_{\rm cutoff}}_\mathsmaller{\frac{n\pi}{d}}  d|{\bf q}| |{\bf q}|^2{e^{-|{\bf q}|\tau}}{}
=\lim_{{\tau\rightarrow 0^+}}\frac{1}{2\pi d}\frac{\partial^2}{\partial \tau^2}{\sum^\infty_{n=0}} \,'\int^{\Lambda_{\rm cutoff}}_\mathsmaller{\frac{n\pi}{d}}  d|{\bf q}| {e^{-|{\bf q}|\tau}}{}\nonumber\\
&=&\lim_{{\tau\rightarrow 0^+}}\frac{1}{2\pi d}\frac{\partial^2}{\partial \tau^2}\frac{1}{\tau}{\sum^\infty_{n=0}} \,'
\left(e^{-\frac{n\pi}{d}\tau}-e^{-{\Lambda_{\rm cutoff}}\tau}\right).\label{intqxqy}
\end{eqnarray}
Apparently, the term of $e^{-{\Lambda_{\rm cutoff}}\tau}$ is being neglected in the first trial, however if $\tau$ is vanishing, then something might go wrong with this term dropped. For a convenience reason, we may choose ${\Lambda_{\rm cutoff}}=1/\tau$ and see if the extra term can be used to cancel against the divergence in eq. (\ref{aveE}). The result from the term of $e^{-{\Lambda_{\rm cutoff}}\tau}$, which is equal to $e^{-1}$ now, is 
\begin{eqnarray}
-\frac{1}{\pi e \tau^3d }\left(\frac{1}{2}+N_{\rm max}\right),
\end{eqnarray}
where $N_{\rm max}$ is the maximal number of the mode number $n$.
The variable $\tau$ now can be made to be 
\begin{eqnarray}
\tau=\frac{3e\,d}{\pi\mathsmaller{\left(\frac{1}{2}+N_{\rm max}\right)}},\label{taucondition}
\end{eqnarray}
so the divergence in the energy density can be canceled and the combined result of $\bar{\varepsilon}=\bar{\varepsilon}_1+\bar{\varepsilon}_2$ is made finite. To seek a deeper reason for  the above condition, the right-hand side  in eq. (\ref{taucondition}) can be regarded as the minimal distance which has physical meanings. Since the mode number of the electromagnetic  waves should not exceed the total number of photons, it gives rise to an upper bound  to the highest energy that a photon can reach, or equivalently we may say that it sets a minimal distance that can be resolved by photons. This accounts for the role of the factor $e^{-|{\bf q}|\tau}$ in eq. (\ref{aveE}), because a photon can not carry energy without any constraint, and the value of $\tau$ sets a characteristic distance for vacuum. The explanation can be illustrated as in Fig. \ref{Fig:cutoffgraph}. The appearance of the divergence is due to a flaw in our original derivation, though the cancellation looks tricky, a finite result  should not be beyond our expectation for there is a non-vanishing $\tau$ in eq. (\ref{aveE}). In fact, when considering the practical setup of the Casimir effect, since the perfect conducting walls do not exist, the photons with a wavelength smaller than the size of a molecular or an atom can not be confined in the space between two plates. It sets a practical maximal number of mode $N_{\rm max}\simeq\frac{d}{a_0}$, where $a_0$ is the Bohr radius, for the $z$-dimension, and accordingly the characteristic length $\tau\simeq \frac{3e}{\pi}a_0$ from eq. (\ref{taucondition}).  Thus, the imperfectness of the conducting walls lowers the maximal number of mode for $a_0$ is much larger than the smallest distance that the natural cutoff, $\Lambda_{\rm cutoff}$, leads to. As a result, the first part, $\bar{\varepsilon}_1$, contributes nothing to the energy density. This is reasonable for there is no difference of the energies between the inside and the outside as the separation of two plates, $d$, is very large.   Now we may turn our focus on the second part, $\bar{\varepsilon}_2$. In the limit of $\tau$ goes to zero, the function $\cosh (|{\bf q}|\tau) n_B(|{\bf q}|)$ can be approximated by $e^{-{\beta_0}|{\bf q}|}$ if ${\beta_0}$ is not a small value. Then the computation of $\bar{\varepsilon}_2$ becomes similar to that of $\bar{\varepsilon}_1$, except that $\tau$ is replaced by the parameter  ${\beta_0}$. The divergence  no longer exists, while instead it becomes a term of $1/{\beta_0}^4$. We obtain
  \begin{eqnarray}
\bar{\varepsilon}_2&=&\lim_{\tau\rightarrow 0^+}\frac{2}{\pi^2}{\sum^\infty_{n=0}} \,'\int^\infty_0 dq_x \int^\infty_0 dq_y|{\bf q}|\cosh (|{\bf q}|\tau) n_B(|{\bf q}|)\nonumber\\&=&\frac{1}{2\pi}\left(\frac{12}{\pi{\beta_0}^4}-\frac{\pi^3}{180\,d^4}+O\left(\frac{1}{d^5}\right)\right).\label{aveE2}
\end{eqnarray}
Therefore, as $d$ goes to cover one of the three dimensions of the whole universe, it is obviously that only the first term in the parenthesis of $\bar{\varepsilon}_2$ survives. The energy density of the electromagnetic vacuum in the universe is 
  \begin{eqnarray}
\bar{\varepsilon}&=&\frac{6}{\pi^2{\beta_0}^4},\label{aveE3}
\end{eqnarray}
which is also a scale invariant quantity after divided by $a^4(\eta_*)$, just like what is obtained in eq. (\ref{dsconst}) from the DeWitt-Schwinger representation. Now we are able to look at the famous discrepancy mentioned in the beginning of this section  and know that, in the imaginary-time field theory, the problem can be conciliated.\par When considering the pressure that induced by the Casimir effect, we may remember that the corresponding Casimir force per unit area, $F(d)$, is obtained from 
\begin{eqnarray*}
F(d)=\left.-\frac{1}{L^2}\frac{d}{ dz}E(z)\right|_{z=d}=-\frac{\pi^2}{240d^4}.
\end{eqnarray*}
 When the distance, $d$, between the two plates is going to be about the size of the universe, the energy $E(d)=(L^2 d) \,\bar{\varepsilon}$ from eq. (\ref{aveE3}), the force per unit area, or say the pressure, $P(d)$, is 
\begin{eqnarray*}
P(d)=\left.-\frac{1}{L^2}\frac{d}{ dz}E(z)\right|_{z=d}=-\bar{\varepsilon}.
\end{eqnarray*}
This is exactly what we expect from a cosmological constant, a negative pressure, which has the same strength as the vacuum energy, and it is derived from the calculation of the Casimir effect in the perspective of the imaginary-time field theory. The origin of the negative pressure would results from the vacuum fluctuations, as illustrated in Figure \ref{Fig:vacind}, a small attractive force is induced between one fermion loop and another induced one in the neighborhood. It is somehow similar to the origin of the van der Waals force \cite{vanderwaals}.
\subsection{Fermion Casimir effect }
\label{fermioncasimir}
In the previous subsection, the energy density and pressure of photons are derived and are found to exactly assume the role of the cosmological constant that we expect from the knowledge of the modern cosmology.  
We would wonder if other types of particles also take parts in the cosmological constant. As for the energy density of fermions, the calculation can be followed from the calculations presented in ref. \cite{huang13b}. The average energy after the summation over the Matsubara frequencies is
\begin{eqnarray}
\langle\mathcal{H}_{\rm D}({\bf p})\rangle
&=&\lim_{\tau\rightarrow 0^+}\frac{1}{\beta}\sum_{n={\rm odd}}\frac{4\xi^2_{\bf p}e^{-i\omega_n\tau}}{(i\omega_n)^2-\xi^2_{\bf p}}
\nonumber\\
&=&\lim_{\tau\rightarrow 0^+}-2\xi_{\bf p}\left(e^{-\xi_{\bf p}\tau}-2\cosh(\xi_{\bf p}\tau)n_F(\xi_{\bf p})\right).\label{hamD}
\end{eqnarray}
Then the derivation is proceeded according to the setup of the Casimir effect. We may also separate the calculation into two parts for the individual terms in the parenthesis of eq. (\ref{hamD}).  After integrating over the momentum $q_x$ and $q_y$, like in eq. (\ref{intqxqy}), the energies per unit area for the respective parts are 
 \begin{eqnarray*}
\hspace{-1cm}\frac{E_{D,1}(d)}{L^2}
=\lim_{\tau\rightarrow 0^+}-\frac{1}{\pi}\frac{\partial^2 }{\partial \tau^2}\frac{1}{\tau}\sum_{n={\rm odd}} e^{-\sqrt{\frac{n^2\pi^2}{4d^2}+m^2}\tau},\hspace{.3cm} \frac{E_{D,2}(d)}{L^2}
=\left.\frac{1}{\pi}\frac{\partial^2 }{\partial \tau^2}\frac{1}{\tau}\sum_{n={\rm odd}} e^{-\sqrt{\frac{n^2\pi^2}{4d^2}+m^2}\tau}\right|_{\tau=\beta_0},
\end{eqnarray*}
where $m$ is the mass of a fermion and the density function $n_{\rm F}(\xi_{\rm p})$ is approximated as $\simeq e^{-\beta_0\xi_{\rm p}}$ for a low temperature. The summation can be further derived by
 \begin{eqnarray}
\hspace{-1cm}\sum_{n={\rm odd}}e^{-\sqrt{\frac{n^2\pi^2}{4d^2}+m^2}\tau}&=&\sum_{n={\rm odd}}e^{-\frac{n\pi\tau}{2d}}-m\tau\sum_{n={\rm odd}}\int^\infty_1dxe^{-\frac{n\pi\tau}{2d}x}\frac{J_1\left(m\tau\sqrt{x^2-1}\right)}{\sqrt{x^2-1}},\hspace{1cm}\label{expmn0}\\
&=&\frac{1}{2\sinh\frac{\pi\tau}{2d}}-\int^\infty_1dx \frac{m\tau}{2\sinh\left(\frac{\pi\tau}{2d}x\right)}\frac{J_1\left(m\tau\sqrt{x^2-1}\right)}{\sqrt{x^2-1}},\label{expmn}
\end{eqnarray}
where $J_\nu(x)$ is the Bessel function of the first kind.  The useful formulas that are used in the eq. (\ref{expmn0}) are 
 \begin{eqnarray}   
&&\int^\infty_1e^{-\alpha x}J_0\left(\beta\sqrt{x^2-1}\right)dx=\frac{1}{\sqrt{\alpha^2+\beta^2}}e^{-\sqrt{\alpha^2+\beta^2}},\label{formula1} \\ &&\hspace{1cm} {\rm and}\,\,
\int^1_0 x^{\nu+1}J_\nu(ax)dx=a^{-1}J_{\nu+1}(a)\,\,{\rm for}\,\,{\rm Re}\,\nu>-1. 
\label{formula2}
\end{eqnarray}
Eq. (\ref{expmn0}) can be obtained by integrating eq. (\ref{formula1}) over the variable, $\beta^2$, then applying eq. (\ref{formula2}) and summing over all odd numbers of $n$.
We shall expand the function $\frac{1}{\sinh(\pi\tau/2d)}$ and $\frac{1}{\sinh(\pi\tau x/2d)}$ in eq. (\ref{expmn}) for $\tau=0^+$, or $\beta_0$, and obtain
 \begin{eqnarray}
&&\hspace{-.5cm}=\frac{1}{2}\left(\frac{2d}{\pi\tau}-\frac{1}{6}\frac{\pi\tau}{2d}+\frac{7}{360}\frac{\pi^3\tau^3}{8d^3}+O(\tau^4)\right)\nonumber\\
&&-\frac{m\tau}{2}\int^\infty_1dx\left(\frac{2d}{\pi\tau x}-\frac{1}{6}\frac{\pi\tau }{2d}x+\frac{7}{360}\frac{\pi^3\tau^3}{8d^3}x^3+O(\tau^4)\right)\frac{J_1\left(m\tau\sqrt{x^2-1}\right)}{\sqrt{x^2-1}}.\hspace{1cm}\label{expansion}
 \end{eqnarray}
As two plates are placed close to each other,  $d\ll 1$, the Casimir effect for the massless fermions, $m=0$, can be recovered
 \begin{eqnarray*}
\frac{E_{D}(d)}{L^2}=-\frac{7}{2880}\frac{\pi^2}{d^3},\,\, {\rm and}\,\,F_D(d)=\left.-\frac{1}{L^2}\frac{d}{ dz}E_D(z)\right|_{z=d}=-\frac{7\pi^2}{960d^4}.
 \end{eqnarray*}
\begin{figure}[t]
\begin{center}
   \subfigure[]{\includegraphics[width=5.8cm]{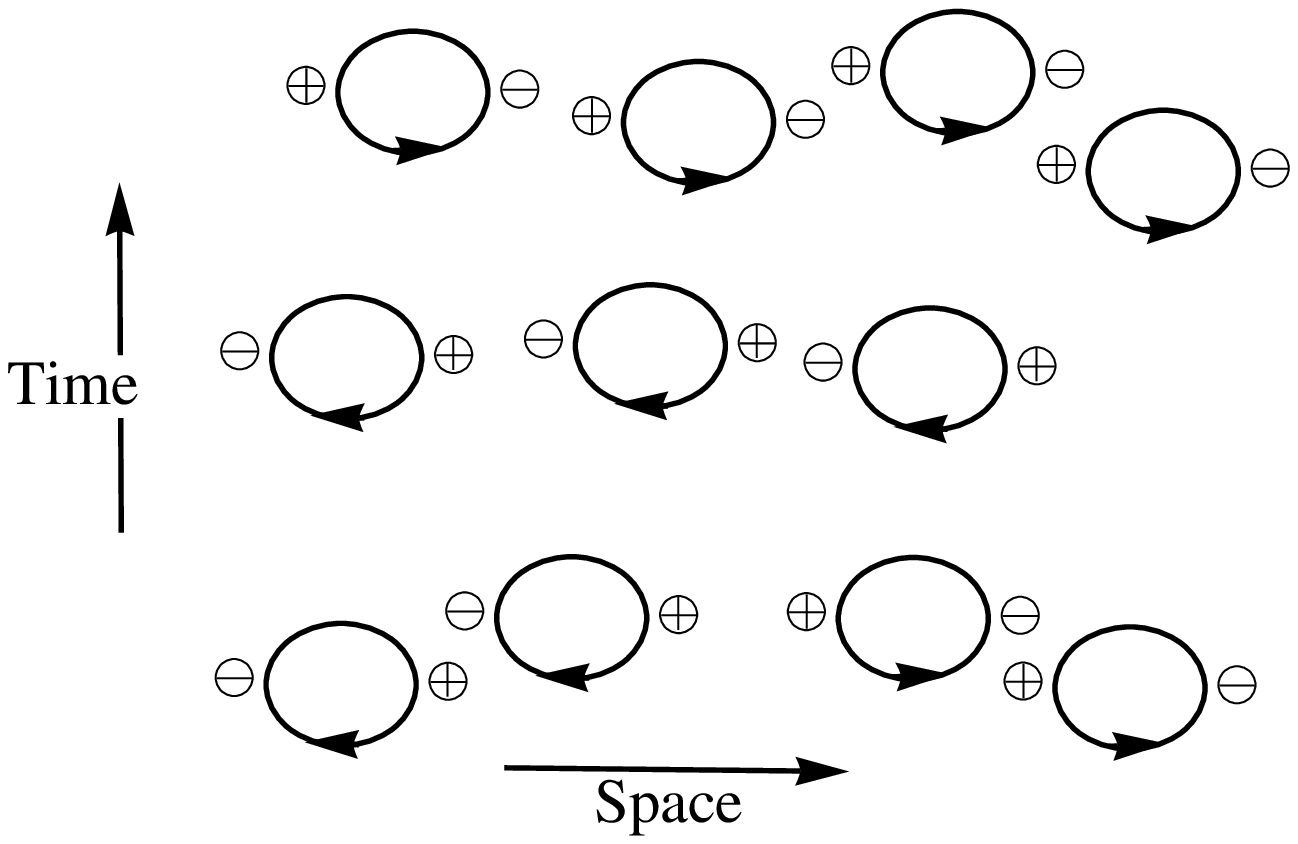}}
   \hspace*{0.01\textwidth}
   \subfigure[]{\includegraphics[width=5.8cm]{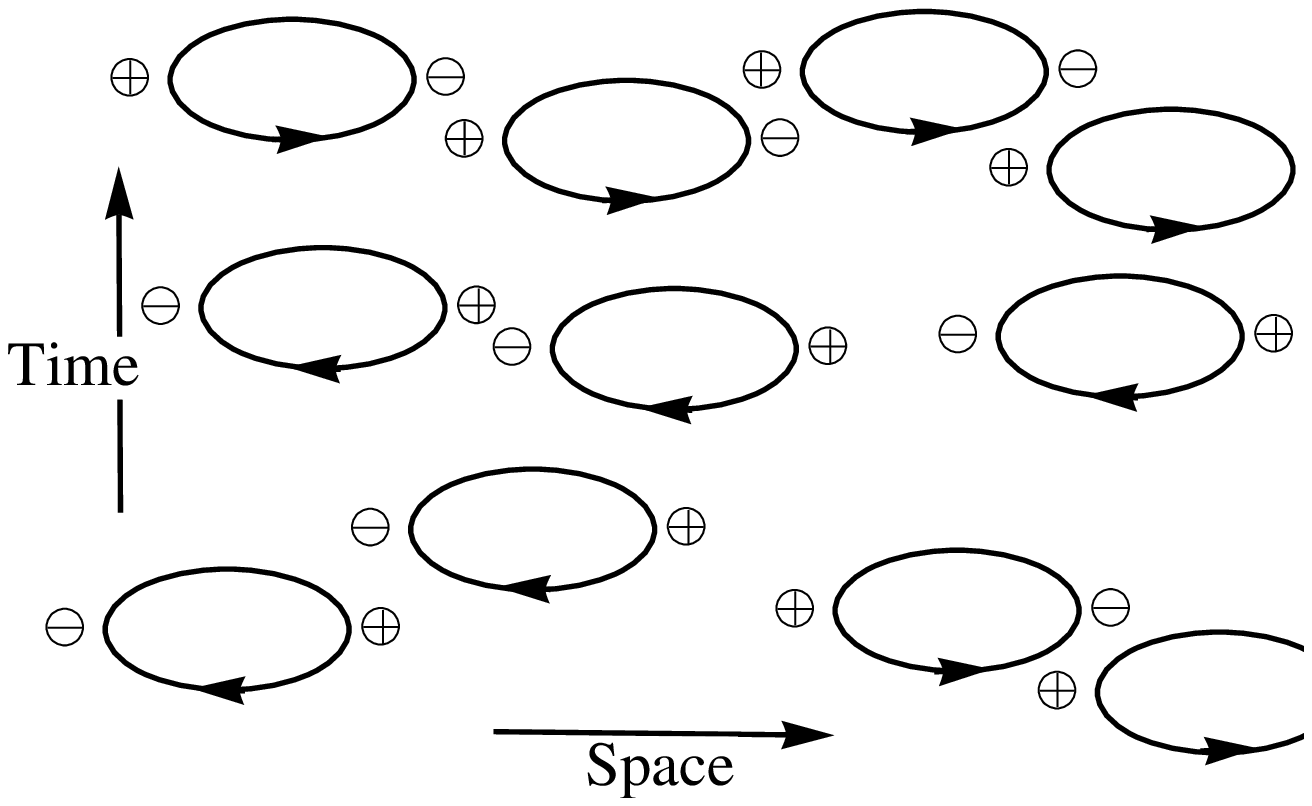}}
   \hspace*{0.0\textwidth}
\caption{\small  The above figure illustrates the induction of vacuum fluctuations for each others in the neighborhood. The loops are the trajectories of charged particles in space-time. Take electrons for example, a virtual electron, $\ominus$, moving along time behaves as negative charged particle and induces one vacuum fluctuation in its side; on the other side of the loop, a virtual positron, $\oplus$, which is moving against time excites another electron loop in the nearby. Microscopically, small attractive forces appear between virtual loops of the electrons; the macroscopic behavior of the system in a very large scale yields a negative pressure in the vacuum. For a higher temperature, the fluctuations are expected to be more violent, as shown in (b), than those of (a) in a lower temperature.  } 
  \label{Fig:vacind}
\end{center}
\end{figure}
What concerns us is if it could also give a finite  prediction  as one of the ingredients for the cosmological constant. The only factors related to the size of the system in eq. (\ref{expmn})  are $\frac{1}{\sinh\left(\frac{\pi\tau}{2d}\right)}$ and  $\frac{1}{\sinh\left(\frac{\pi\tau x}{2d}\right)}$; they can be approximated by the $\frac{1}{\frac{\pi\tau}{2d}}$ and 
$\frac{1}{\frac{\pi\tau x}{2d}}$ if $d$ is about the size of the universe. Similar to the case of electromagnetic waves, for the first part of the energy per unit area $\tau$ is taken to an infinitesimal value, the divergences  caused by the first term in the factor  $\frac{1}{\sinh\left(\frac{\pi\tau}{2d}\right)}$ in eq. (\ref{expansion}) can be canceled by  a chosen cutoff for momentum  $q_x$ and $q_y$. For the integral of the factor $\frac{1}{\sinh\left(\frac{\pi\tau x}{2d}\right)}$, consider the leading term of the factor in the expansion and change the variable to $y=m\tau \sqrt{x^2-1}$, the integral becomes
 \begin{eqnarray*}
\frac{m d}{\pi }\int^\infty_1dx\frac{J_1(m\tau\sqrt{x^2-1})}{x\sqrt{x^2-1}}=\frac{m^2 \tau d}{\pi }\int^\infty_0dy\frac{J_1(y)}{y^2+m^2\tau^2}=\frac{1-m\tau K_1(m\tau)}{\pi\tau}d.
 \end{eqnarray*}
The Bessel function $K_1(x)\simeq \frac{1}{x}+O(x)$ near the point $x= 0$, therefore the integral of $x$ in eq. (\ref{expansion}) makes no contribution to the energy density for $\tau\rightarrow 0^+$ either. The only term in eq. (\ref{expansion}) could possibly contribute in this limit of $\tau$ is the one proportional to $\tau^3$, however as $d$ goes to infinity, the contribution also diminishes. Therefore only the second part for $\tau=\beta_0$ could contribute to the cosmological constant. The volume between two plates is $L^2 d$. In the limit of $m\beta_0\gg 1$, the energy density and the pressure are 
 \begin{eqnarray*}
&&\hspace{-1cm}\bar{\varepsilon}_D=\frac{E_D(d)}{L^2 d}=
\frac{e^{-m\beta_0}}{\beta_0^4}\left(\frac{(m\beta_0)^\frac{5}{2}}{\sqrt{2}\pi^\frac{3}{2} }+\frac{27(m\beta_0)^\frac{3}{2}}{8\sqrt{2}\pi^\frac{3}{2} }+\frac{705(m\beta_0)^\frac{1}{2}}{128\sqrt{2}\pi^\frac{3}{2} }+O(\frac{1}{(m\beta_0)^\frac{1}{2}})\right),\nonumber\\
&&\,\,{\rm and}\hspace{.5cm}
P_D(d)=\left.-\frac{1}{L^2}\frac{d}{ dz}E_D(z)\right|_{z=d}=-\bar{\varepsilon}_D.
 \end{eqnarray*}
The energy density and pressure is strongly suppressed by the exponential factor $e^{-m\beta}$, so   massive fermions do not make a substantial contribution to the cosmological constant. 
 For example, the value, $m\beta$, for an electron at a temperature $1$K is about $\sim10^{10}$, not to mention that a black hole of one solar mass has a temperature $\sim 10^{-7}$ K. For the case of massless fermions, like the neutrinos in the Standard Model, the contribution from one species  is 
 \begin{eqnarray*}
\bar{\varepsilon}_D=\frac{6}{\pi^2\beta_0^4},\,\,{\rm and}\,\,
P_D(d)=\left.-\frac{1}{L^2}\frac{d}{ dz}E_D(z)\right|_{z=d}=-\bar{\varepsilon}_D.
 \end{eqnarray*}
The results are the same as those of the electromagnetic waves. In Figure \ref{Fig:vacind}, it explains how a negative pressure is generated due to the induction of the vacuum fluctuations.







\section{Conclusion}
\label{conclusion}
In this paper, I have discussed two possible origins of the cosmological constant  from the point of view of the imaginary-time field theory and incorporated the concept of the thermal time with scale invariance of the actions in the theory.
In the first approach, the DeWitt-Schwinger representation, which used to predict a divergent value from the field theory, is applied. The Green function of the electromagnetic waves adopted from the imaginary-time field theory, is used  in a similar manner and gives up a finite result for the cosmological constant. The obtained energy density is proportional to $T^4$, similar to that  of the black-body radiation. From the concept of the thermal time and scale invariance, the scale factor in the FRW metric is found to be determined by the temperature, $T$, and the temperature decreases as the universe expands. This gives rise to a constant energy density over space during the expansion of the universe.  The value of the cosmological constant then corresponds to a specific temperature that might happen sometime due to some special event in the early evolution of the universe. Like the case of the CMB, it imprinted the temperature about $3000K$ for the beginning of the recombination era, when the universe was about $37900$ years old, and afterwards cooled down to the current observed value $2.725$ K.  In the second method, the energy density and the pressure of vacuum are computed through the way of the Casimir effect. From the old estimation of the vacuum energy, the summation over all of the zero-energy mode numbers results in a big discrepancy between the theoretical and experimental values. The hamiltonian of the imaginary-time  field theory in calculating the Casimir vacuum energy and pressure gives a satisfying prediction for the cosmological constant, including a constant energy density and a negative pressure with an equation of state coefficient $w=-1$, and remedies the mysterious discrepancy. An important requirement that is obtained from both of the approaches is that the number of mode of a field can not exceed the number of the particles in the vacuum. This criteria sets natural cutoffs  on the phase space of the energy and the momentum, and is required to remove  redundant divergences  in the derivation.




%
%

\end{document}